\documentclass[twocolumn,numberedappendix,tighten]{aastex61}
\usepackage{amsmath,amssymb,amstext}
\usepackage[all]{hypcap} 
\usepackage{amssymb,amsmath,amsthm}
\usepackage{longtable}
\usepackage{amssymb}
\usepackage{amsmath, xspace}
\usepackage{graphics,graphicx} 
\usepackage{rotating}
\usepackage{color}
\usepackage{tablefootnote}
\newcommand{\xmm}{{\it XMM-Newton}}
\newcommand{\chandra}{{\it Chandra}}

\newcommand{\arxvi}{Ar~{\sc xvi}}
\newcommand{\arxvii}{Ar~{\sc xvii}}

\newcommand{\adb}{AtomDB~v3.0.8}
\newcommand{\adbold}{AtomDB~v2.0.2}
\def\ionch#1#2{#1$^{#2+}$}
\newcommand{\ardr}{Ar {\sc xvii} He$\beta$ DR}

\shortauthors{Bulbul et al.}

\begin{document}

\title{Laboratory Measurements of  X-Ray Emission from Highly Charged Argon Ions}

\author{Esra Bulbul}
\affiliation{Massachusetts Institute of Technology, Kavli Institute for Astrophysics \& Space Research, 77 Massachusetts Ave, Cambridge, MA 02139, USA}
\affiliation{Harvard-Smithsonian Center for Astrophysics, 60 Garden Street, Cambridge, MA, 02138, USA}

\author{Adam Foster}
\affiliation{Harvard-Smithsonian Center for Astrophysics, 60 Garden Street, Cambridge, MA, 02138, USA}

\author{Gregory V. Brown}
\affiliation{Lawrence Livermore National Laboratory, 7000 East Avenue, Livermore, CA, 94550, USA}

\author{Mark W. Bautz}
\affiliation{Massachusetts Institute of Technology, Kavli Institute for Astrophysics \& Space Research, 77 Massachusetts Ave, Cambridge, MA 02139, USA}

\author{Peter Beiersdorfer}
\affiliation{Lawrence Livermore National Laboratory, 7000 East Avenue, Livermore, CA, 94550, USA}

\author{Natalie Hell}
\affiliation{Lawrence Livermore National Laboratory, 7000 East Avenue, Livermore, CA, 94550, USA}

\author{Caroline Kilbourne}
\affiliation{NASA Goddard Space Flight Center, 8800 Greenbelt Road, Greenbelt, MD, 20771, USA}

\author{Ralph Kraft}
\affiliation{Harvard-Smithsonian Center for Astrophysics, 60 Garden Street, Cambridge, MA, 02138, USA}

\author{Richard Kelley}
\affiliation{NASA Goddard Space Flight Center, 8800 Greenbelt Road, Greenbelt, MD, 20771, USA}

\author{Maurice A. Leutenegger}
\affiliation{NASA Goddard Space Flight Center, 8800 Greenbelt Road, Greenbelt, MD, 20771, USA}
\affiliation{Department of Physics, University of Maryland Baltimore County, 1000 Hilltop Circle, Baltimore, Maryland 21250, USA}

\author{Eric D. Miller}
\affiliation{Massachusetts Institute of Technology, Kavli Institute for Astrophysics \& Space Research, 77 Massachusetts Ave, Cambridge, MA 02139, USA}

\author{F. Scott Porter}
\affiliation{NASA Goddard Space Flight Center, 8800 Greenbelt Road, Greenbelt, MD, 20771, USA}

\author{Randall K. Smith}
\affiliation{Harvard-Smithsonian Center for Astrophysics, 60 Garden Street, Cambridge, MA, 02138, USA}

\correspondingauthor{Esra~Bulbul}
\email{ebulbul@cfa.harvard.edu}

\begin{abstract}

Uncertainties in atomic models will introduce noticeable additional systematics in calculating the flux of weak dielectronic recombination (DR) satellite lines, affecting the detection and flux measurements of other weak spectral lines.  One important example is the \ardr, which is expected to be present in emission from the hot intracluster medium (ICM) of galaxy clusters and could impact measurements of the flux of the 3.5~keV line that has been suggested as a secondary emission from a dark matter interaction. We perform a set of experiments using the Lawrence Livermore National Laboratory's electron beam ion trap (EBIT-I) and the X-Ray Spectrometer quantum calorimeter (XRS/EBIT), to test the \ardr\ origin of the 3.5~keV line. We measured the X-ray emission following resonant DR onto helium-like and lithium-like Argon using EBIT-I's Maxwellian simulator mode at a simulated electron temperature of $T_{e}$=1.74~keV. The measured flux of the \ardr\ lined is too weak to account for the flux in the 3.5~keV line assuming reasonable plasma parameters. We, therefore, rule out \ardr\ as a significant contributor to the 3.5~keV line. A comprehensive comparison between the atomic theory and the EBIT experiment results is also provided.

\end{abstract}


%
\section{Introduction}\label{sec:intro}
Since its discovery, the nature of dark matter has been one of the prime problems of physics. A range of exotic particles, which could constitute the dark matter content of the Universe, have been widely investigated by ground- and space-based direct and indirect searches. Observations of dark matter dominated objects with space telescopes provide an avenue for indirect detection of secondary emission from dark matter interactions. An intriguing detection of an emission line at $\sim$ 3.5 keV has been reported by \citet[][B14a hereafter]{b14a} in the stacked observations of clusters of galaxies and in many other dark matter-dominated-objects \citep{bo14, bo15, urban15, franse16, neronov16, bulbul16, cappelluti17}. The origin of the emission line is still under discussion, with one proposal being that the X-rays are produced by the secondary emission of dark matter particles \citep[e.g.][]{abazajian17,conlon16}. Alternative astrophysical origins of the unidentified line have been extensively discussed in Bu14a and \citet{gu2015}, one of which is the possibility that it results from an unexpectedly strong dielectronic recombination (DR) satellite line created when He-like \ionch{Ar}{16} recombines to form Li-like \ionch{Ar}{15}. Uncertainties on the atomic calculations of weak satellite lines, such as the \ardr\ transition  $1s\ 2p\ 3p\  {}^2D_{5/2} \rightarrow 1s^2 2p\ {}^2P_{3/2}$, may introduce unknown systematics on the measured flux of the 3.5~keV line and provide an alternative explanation of the origin of the line \citep[][B14a, B14b hereafter]{b14a, b14b}.

Measurements performed using the EBIT-I electron beam ion trap \citep[EBIT;][]{Levine88, Marrs88, Marrs08, bei2003, Bei2008} located at Lawrence Livermore National Laboratory (LLNL) provide an avenue for testing the \ardr\ origin of the 3.5 keV line and evaluation of the atomic databases, e.g. \adbold\ and FAC \citep{foster2012}. Using a unique operating mode, EBIT-I generates a quasi-Maxwellian distribution of electron energies \citep{savin2000a,savin2008a} at electron temperatures similar to those found in astrophysical objects in thermal equilibrium. Measurements using this mode have been used to test atomic theory for highly charged ions of iron \citep{gu2012a} and gold \citep{may2004b, may2005a}. In addition, the accuracy of the Maxwellian generator has also been tested using well-known intensity ratios from helium-like ions \citep{savin2000a, savin2008a} and hydrogen-like ions \citep{gu2012a}. Measurements using this mode include all relevant population processes in a complete way, i.e., there is no limit to the number of transitions that are included, unlike in computer models.  We have used this mode to generate an argon plasma at $kT_{e}$~$\sim$~1.7~keV and to measure the emission using a high-resolution quantum calorimeter, which is similar to the Soft X-ray Spectrometer (SXS) \citep{kelley16} flown on the {\it Hitomi} X-ray Observatory \citep{Takahashi16} and which includes all line emission from K-shell transitions in highly charged argon spanning the 3~keV to 4~keV band. These measurements include emission from the \ion{Ar}{17} He-$\alpha$ complex \ion{Ar}{18}~Ly-$\alpha$, and \ion{Ar}{17}~He-$\beta, \gamma, \delta$ and $\epsilon$ lines, with their relative intensities to the He-like \ion{Ar}{17} He-$\alpha$ resonance line at 3.12~keV. These new measurements are complementary to measurements conducted earlier using a high resolution crystal spectrometer attached to the Princeton Large Torus \citep{bei95}. In that work, the authors measured the $K\beta$ line of He-like \ionch{Ar}{16} in the energy range of 3.54~keV to 3.71~keV and successfully identified \arxvi\ lines produced by dielectronic recombination from Ar$^{16+}$ to Ar$^{15+}$ as well as lines from electron-impact excitation, the latter not only from He-like Ar$^{16+}$, but also from lithium-like Ar$^{15+}$ and beryllium-like Ar$^{14+}$. Here, we use EBIT-I to directly measure the spectrum of highly charged argon ions in a simulated Maxwellian plasma (see Section~\ref{sec:ExpDes}) reaching down to 3~keV. Using the data from both the new measurements and from those of \cite{bei95}, a stringent test of the models used by B14a is completed. We compare the line ratios measured by the new EBIT measurements with the observed line ratios in \xmm\ and \chandra\ observations of stacked clusters where the unidentified line has been observed with the highest significance.

This paper is organized as follows. We provide details of the EBIT-I experiment design in Section \ref{sec:ExpDes}. In Section \ref{sec:AtomicModel}, we provided details of the atomic models used: since the older version of AtomDB v2.0.2 used in B14a cannot analyze the EBIT-I plasma, we include here a comparison between the AtomDB versions. The analysis in Section \ref{sec:results} compares the EBIT-I spectrum with the FAC and \adb\ models, then explores the effects of these results on the discoveries in B14a.
We finally summarize our conclusion in Section \ref{sec:conc}. All errors quoted throughout the paper correspond to 68\% single-parameter confidence intervals.

\begin{table*}
\caption{The atomic data sources for Argon in AtomDB v2.0.2 and AtomDB v3.0.8. CH = Chianti v7, \citep{2009A+A...498..915D}. D = \citet{1988CaJPh..66..586D}. W = \citet{Whiteford2001}. VS = \citet{1980ADNDT..25..311V}. L = \citet{2011A+A...528A..69L}. LI= \citet{2015A+A...583A..82L}. E = \citet{1977JPCRD...6..831E}. N = \citet{NISTASD5.3}. P = \citet{2008ApJS..177..408P}. FAC=Flexible Atomic Code, similar to this work. AS = Autostructure \citep{Badnell}\label{tab:sources}}
\centering
\begin{tabular}{llllll}
\hline \hline
Ion & Wavelengths & Einstein A & Coll. Exc. & DR Satel. & Inner-shell\\
\hline
\multicolumn{3}{l}{AtomDB v2.0.2}\\
\ionch{Ar}{14}&  CH & CH & CH & None & None\\
\ionch{Ar}{15}&  CH & CH & CH&  None& None\\
\ionch{Ar}{16}&  D, W& W & W           &  VS& None\\
\ionch{Ar}{17}&  E, AS & AS     & FAC           &  VS& None\\
\hline
\multicolumn{3}{l}{AtomDB v3.0.8}\\
\ionch{Ar}{14}&  CH & CH & CH & None & P\\
\ionch{Ar}{15}&  N & L & CH &  None & P\\
\ionch{Ar}{16}&  D, W & W & W           &  VS & FAC\\
\ionch{Ar}{17}&  E, N & FAC    & LI &  VS & None\\
\hline
\end{tabular}
\vspace{4mm}
\end{table*}
\section{Experiment Design}
\label{sec:ExpDes}

The results presented here are measured using EBIT-I's Maxwellian simulation mode \citep{savin2000a,savin2008a}. In this mode, the electron beam is ``swept" in a voltage pattern that, when averaged over several cycles, corresponds to a Maxwell-Boltzmann electron distribution at an experimentally controlled temperature. 
The electron beam energy is ``swept" from below threshold for any relevant process that may contribute to the line emission of interest, such as low-energy DR resonances, up to electron energies $\geq$ 6 times the simulated temperature. For example, for the $\sim$2~keV plasma presented here where the focus is on K-shell emission from highly charged argon ions and where all the relevant dielectronic resonance energies are above 2~keV, the beam is swept from 1.5 keV up to 24~keV. The electron density for this measurement is $\sim$ $10^{11}$ cm$^{-3}$. One caveat of the Maxwellian mode is that the plasma generated in EBIT has a lower average charge than a plasma in true thermal equilibrium, thus  intensities of line emission from different ions will not be representative of a Maxwellian plasma.  However, the relative intensities of emission lines whose excited states are dominated by processes involving a single ion, such as ratios of two He-like lines or of a resonance line and a DR satellite from the same parent charge state, are the same as found in a true Maxwell-Boltzmann plasma \citep{savin2000a,savin2008a,gu2012a}. This is supported by the results presented here for both He-like argon and H-like argon. 

The spectra are measured using the X-Ray Spectrometer (XRS)/EBIT quantum calorimeter \citep{porterfs2000a,porter2004a} designed and built at the NASA Goddard Space Flight Center. The XRS/EBIT is an energy-dispersive spectrometer with an energy resolution of $\Delta \ E \sim \ 5$~eV, a bandwidth that spans the range from below 500~eV to about 10~keV, and a quantum efficiency of 100\% for photon energies up to $\sim$6~keV. Owing to the fact that the XRS/EBIT calorimeter array operates at 60 mK, it must be shielded from higher-temperature sections of the spectrometer. This is achieved via four aluminized polyimide blocking filters. In addition, there is an insertable aluminized polyimide filter mounted between EBIT-I and the XRS/EBIT. This filter was used to reduce flux from low energy photons. The combined thickness of the filters is 3824 \AA\ of aluminum and 14310 \AA\ of polyimide. The relative X-ray transmission efficiency through these filters for the photon energies studied here is 3\%.
The XRS/EBIT  viewed the trapped ions through one of EBIT-I's six radial ports oriented at 90$^{\circ}$ to the electron beam. Neutral argon gas is introduced to the trap continuously via a differentially pumped, collimated ballistic gas injector, also attached to one of the six radial ports. Once injected, argon atoms intersect the beam and are then ionized and trapped.

\section{Atomic Modeling}
\label{sec:AtomicModel}

\subsection{AtomDB}
\label{sec:AtomicModel:AtomDB}

The AtomDB project couples an atomic database with a model of optically thin, collisionally ionized plasma to produce emission spectra predictions for astrophysical plasma. This model is known as the \textit{apec} model in the XSPEC analysis suite \citep{arnaud1996}, and was used in Bu14a and Bu14b. These papers used \adbold\ models \citep{foster2012}, while in this work we use the more recent \adb. The atomic data is largely similar between the two, but some fundamental changes to the structure of the database and code have made this later version more suitable to this work.

The AtomDB atomic database consists of data from a wide range of sources for the emissivities. For Argon, the ionization and recombination rates are taken from \citet{Bryans2006, Bryans2009}. For each ion for \ionch{Ar}{14} to \ionch{Ar}{17}, we list the sources of other atomic data in Table \ref{tab:sources}. Note that for each process within each ion, sometimes several different data sources are used based on the coverage of the published data sets. Therefore, in Table \ref{tab:sources} we list only those most relevant to the ions and lines observed in this work. All of the atomic data is available online \footnote{http://www.atomdb.org/}.

The updates to both the atomic data and the underlying structure of AtomDB between version 2.0.2 and 3.0.8 are minor. The changes are adjusted wavelengths to NIST values where possible: this has little effect on this work as typical adjustments were on the 1-2 eV range, significantly smaller than the spectral resolution of the instruments in B14a.
The more significant update is to include representation of inner shell excitation, and generally improve representation of non-equilibrium plasma. This involves using published data for inner shell ionization and fluorescence yields, as well as new Li-like electron collision data to include the effect of the inner shell excitation. In addition, the plasma calculations are reworked completely. Instead of calculating the emission based upon the ion containing the transition (so for the He$\alpha$ line, looking at \ionch{Ar}{16}), the emissivities are broken up by parent ion.
Thus an ion of, say, \ionch{Ar}{16} can give rise to lines from \ion{Ar}{16} due to dielectronic or radiative recombination, \ion{Ar}{17} due to excitation and \ion{Ar}{18} due to excitation-autoionization or direct ionization. Since emissivity is calculated by ion, it is possible to model the ionization balance found in the EBIT-I's Maxwellian simulation mode.

DR Satellite lines are included in AtomDB from \citet{1980ADNDT..25..311V}. This data set includes
capture from H- and He-like Ar ions, but not Li-like. Due to the enhanced Li-like DR line presence, we have added the $1s2s2l3l'$ satellite lines to our model, using the parameters from Table VI of \cite{bei95}. In addition, the $1s3l3l'$ satellites from capture from He-like ions is also not in \adb, so the data from Table V of that paper has also been included. All of these lines lie between 3.54 and 3.69~keV, and therefore could be an additional source of emissivity missing from the models in B14a. These satellite lines, calculated by HULLAC \citep{bar2001}, have been verified experimentally by \cite{bei95} in similar plasma temperatures to those expected in clusters, i.e. $kT=2.3$keV. 

\begin{table*}[ht!]
\centering
\caption{Integrated Normalized Flux of Ar lines in a 1.74~keV Maxwellian plasma from \adbold\ and \adb. Differences to Table \ref{tab:summed} are due to the differences in charge balance between the fitted model and the theoretical Maxwellian.}
\scalebox{0.9}{
{\renewcommand{\arraystretch}{1.0}
{ \normalsize \begin{tabular}{lcccccc}
\hline \hline
Line		&	Energy (keV)	& AtomDB v2.0.2 & AtomDB v3.0.8 & Ratio	\\
\hline
Ar {\sc XVII} He$\alpha$				& 3.080 -- 3.150 & 11.404 & 11.420 & 1.00\\
Ar {\sc XVIII} Ly$\alpha$ DR			& 3.260 -- 3.315 &  0.378 &  0.384 & 0.98 \\
Ar {\sc XVIII} Ly$\alpha$				& 3.315 -- 3.350 &  2.433 &  2.424 & 1.00 \\
Ar {\sc XVII} He$\beta$ DR	    	    & 3.600 -- 3.645 &  0.100 &  0.097 & 1.03 \\
Ar {\sc XVII} He$\beta$				& 3.670 -- 3.700 &  0.849 &  0.832 & 1.02 \\
Ar {\sc XVII} He$\gamma$				& 3.850 -- 3.900 &  0.293 &  0.287 & 1.02 \\
Ar {\sc XVIII} Ly$\beta$			& 3.925 -- 3.945 &  0.318 &  0.298 & 1.07 \\
Ar {\sc XVII} He$\delta$				& 3.950 -- 3.975 &  0.126 &  0.123 & 1.02 \\
Ar {\sc XVII} He$\epsilon - \iota$     & 4.000 -- 4.082 &  0.168 &  0.176 & 0.96 \\
\hline\hline
\end{tabular}}}}
\label{tab:atomdbversion}
\vspace{4mm}
\end{table*}

\subsubsection{Comparison of AtomDB 3.0.8 to AtomDB 2.0.2}
It is not possible to analyze the EBIT-I data with \adbold\ due to the structural differences outlines above. We have therefore analyzed the data in this paper with \adb, and in this section we  explore the differences between \adbold\ and \adb\ to show that results obtained in this work are similar to what would have happened in B14a.

The effect of changing between these two data sets is shown in Table \ref{tab:atomdbversion} for a Maxwellian plasma with kT=1.74~keV(see Section \ref{sec:results} for the details on the temperature estimate). The emissivity of the strongest lines changes by less than 5\%, so any changes due to this are negligible compared to the factor of 2 difference between the experiment and the AtomDB values. We shall ignore any contributions from this for the remainder of this work.

\subsection{The Flexible Atomic Code (FAC)}

The Flexible Atomic Code \citep[FAC;][]{gu2004,gu2008} is a comprehensive code that calculates energy levels, transition rates and cross sections. Additionally, FAC contains a collisional radiative model that estimates the line strengths for optically thin plasmas based on the atomic physics parameters previously calculated with FAC. The ion abundances, choice of processes to include, electron density, and electron distribution are user input parameters. The electron energy distribution can be chosen freely from pre-defined functions, e.g., a Gaussian distribution for electron beams or a Maxwellian distribution for thermal plasmas, or user-defined through a table model. 

Here, we used FAC to calculate the energy levels for H-like through B-like Ar for configurations of the form: $n\ell$ (H-like); $1s^2$, $1s\,n\ell$, $2\ell\,2\ell\prime$, and $2\ell\,n\ell\prime$ (He-like); $1s^2\,2\ell^x$, $1s^2\,2\ell^{x-1}\,n\ell\prime$, $1s\,2\ell^y$, $1s\,2\ell^{y-1}\,n\ell\prime$  (Li- to B-like), allowing for spectator electrons up to n=5. Transition rates and cross sections were calculated for transitions between any of these configuration groups. We then used the data to calculate the plasma model at temperatures in the range 1.2 to 3.0 keV, including radiative decay, collisional (de-)excitation and ionization, radiative recombination, dielectronic recombination, and autoionization. For the relative ion abundances we used FAC's default charge balance at the respective temperatures. The Maxwellian electron energy distribution for each temperature was integrated between 50\,eV and 20\,keV and an electron density of $1\times10^{11}\,\mathrm{cm}^{-3}$ was assumed. We then convolved the resulting line strengths for each transition with a Gaussian line profile of 5.0\,eV FWHM to produce the final spectrum. 

To obtain the theoretical DR/Ly$\alpha$ line ratio, we then summed the spectra in the range of 3250--3307\,eV for the DR lines and 3307--3340\,eV for the Ly$\alpha$ lines such that the unresolved DR resonance blending with the Ly$\alpha$ lines are included in the line ratio. Neglecting the unresolved blends to Ly$\alpha$ would falsely result in  an increase of the ratio by between 23\% at 1.2\,keV and 6\% at 3.0\,keV electron temperature. 
\section{Results}
\label{sec:results}

\begin{figure}[h!]
\centering
\includegraphics[width=0.48\textwidth]{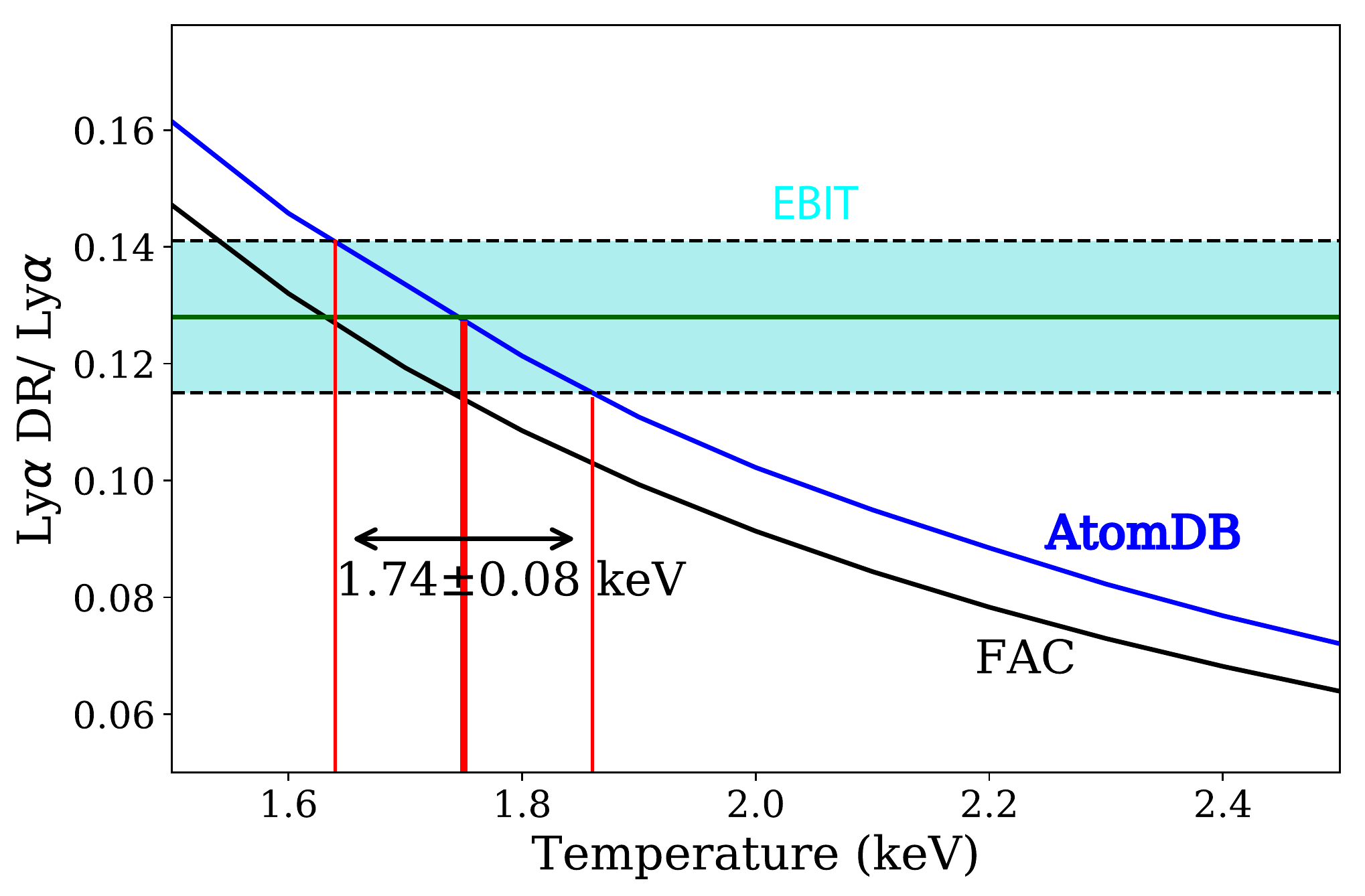}
\caption{\adb\ and FAC predicted temperature dependence of the \ion{Ar}{18}~Ly$\alpha$ DR to the \ion{Ar}{18}~Ly$\alpha$\ lines are shown. The line ratio is compared with the EBIT measurement with its uncertainty limits (shaded cyan region). \adb\ predictions show that the EBIT measurements are consistent with a Maxwellian $T_e = 1.74\pm0.08$~keV. }
\label{fig:kT}
\end{figure}

The electron temperature of the EBIT-I plasma is determined using the line ratio of the \ion{Ar}{18}~Ly$\alpha$ DR lines to their parent \ion{Ar}{18}\ Ly$\alpha$ line \citep{gu2012a}. This method has been well tested and only relies on the presence of hydrogenic ions.  Figure \ref{fig:kT}  shows the temperature dependence of this ratio given by \adb\ and FAC. Before comparing the measured ratio to theory, polarization effects must be taken into account. As a result of the unidirectional electron beam, electron beam ion traps produce non-isotropic, polarized radiation which depends on both electron impact energy and transition \citep{beiersdorfer1996c,gu1999}.  This effect is reduced by the fact that electrons in the beam have a spiral trajectory caused by their thermal velocity when they are generated at the electron gun. In general, polarization is reduced as a function of electron impact energy. 
The polarization of Ly${\alpha1}$ is measured to be $\sim$ 0.1 by \citet{nakamura01}. This result is consistent with a more recent theory \citep{Bostock2009}.  
In the case of the DR emission, a range of polarization values are predicted owing to the fact that several transitions of the type $1s 2\ell$ make up this feature make up this feature. The DR features consist of several transitions with a range of polarization values, most of which are positive. Using the calculations of \citep{chen95} as a guide, we assign an average value  of $P= 0.5$ to all the H-like Ar DR satellites.
Because of the depolarizing effect of the spiraling beam and the fact that the electron beam energy is swept across a large range for these measurements, we make no correction for polarization for this ratio and assign an uncertainty resulting from polarization in the ratio of $\pm$10\%. This is determined by nearly the full range of polarization correction factors, including the possibility of the DR polarization being largely negative. This uncertainty, combined with the statistical uncertainty of $\sim$ 7\%, gives a ratio of the \ion{Ar}{18}\  Ly$\alpha$ DR line to the parent line of 0.135 $\pm$ 0.016. This ratio corresponds to a plasma temperature of $1.74\pm0.08$~keV based on the AtomDB model and 1.64$\pm$0.08~keV based on the FAC model (see Figure \ref{fig:kT}). 

Using the Maxwellian temperature of $kT_e= 1.74$~keV, we calculate the spectra created by \ionch{Ar}{14} -- \ionch{Ar}{18} using \adb. The lines are broadened with a FWHM of 5~eV to match the observed EBIT calorimeter response. In a true Maxwellian plasma, such as is expected in central region of galaxy clusters, the charge state distribution for a $kT_e=1.74$~keV plasma would be multiplied by each ion's emissivity, i.e.
\begin{equation}
\mbox{Counts}_{tot}(E) = C\sum^{z1} \varepsilon_{z1}(E) N_{z1}
\end{equation}

\noindent where $C$ is a constant coefficient incorporating the volume, number density of electrons and argon ions, and detector efficiency, $\varepsilon(E)$ is the emissivity in each energy bin, $E$, and $N_{z1}$ is the fraction of argon ions in ion state $z1$, where $z1$ is the ion charge +1.

\begin{table}
\caption{The ion fraction ($C_{z1}$) obtained from a theoretical Maxwellian and for fits to the EBIT spectrum.\label{tab:ionfrac}}
\centering
\begin{tabular}{llll}
\hline \hline
Ion & Maxwellian & EBIT & EBIT \\
    &            & (AtomDB) & (FAC)\\
\hline
\ionch{Ar}{13} (B-like) & 0.000 & 0.000 & 0.017\\
\ionch{Ar}{14} (Be-like) & 0.001 & 0.087 & 0.087\\
\ionch{Ar}{15} (Li-like) & 0.024 & 0.144 & 0.289\\
\ionch{Ar}{16} (He-like) & 0.577 & 0.681 & 0.513\\
\ionch{Ar}{17} (H-like) & 0.330 & 0.086 & 0.094\\
\ionch{Ar}{18} (fully stripped) & 0.068 & 0.001 & 0.001\\
\hline
\end{tabular}
\end{table}

As we do not have a ionization equilibrium charge state distribution in a plasma with constituents having Maxwellian velocity distributions in the measurement, we determine the ion charge states by treating each ion as having an independent scaling factor, i.e.

\begin{equation}
\mbox{Counts}_{tot}(E) = \sum^{z1} C_{z1} \varepsilon_{z1}(E) N_{z1}
\end{equation}
\noindent and then fitting the data in the 3--4~keV band to these spectra to obtain the $C_{z1}$. The ratio of these determines the charge state distribution. We obtain best-fit ion fractions as shown in Table \ref{tab:ionfrac}, showing a significant increase in the Li-like Ar fraction in the EBIT at the expense of the H-like stage compared with a Maxwellian plasma.

\begin{figure}
\centering
\includegraphics[width=0.47\textwidth]{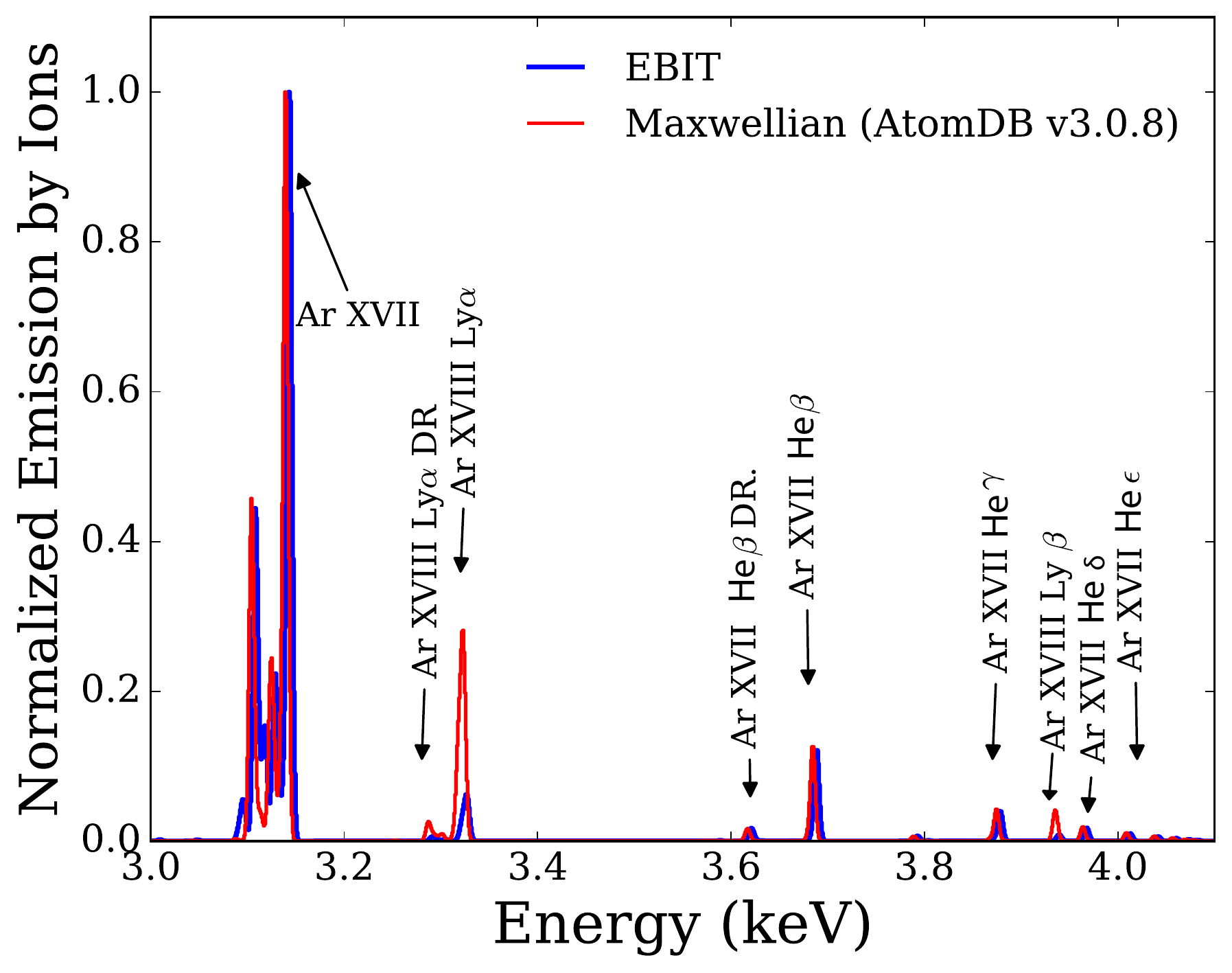}
\caption{A comparison of the Ar emission spectrum obtained from the EBIT and the modeled Maxwellian plasma obtained from \adb\ in the full 3--4~keV energy band. The normalized y-axis is normalized to the \ion{Ar}{17} resonance line. The EBIT spectrum is shifted by +5~eV to show the difference between \adb\ and EBIT spectra. The emission lines from different Ar species are marked on the spectrum. Large discrepancies in the \ion{Ar}{18} lines are due to the non-Maxwellian equilibrium ionization balance in the EBIT, which is assumed in the AtomDB spectrum.
\label{fig:fullspec}}
\end{figure}

\begin{figure*}
\centering
\includegraphics[width=1\textwidth]{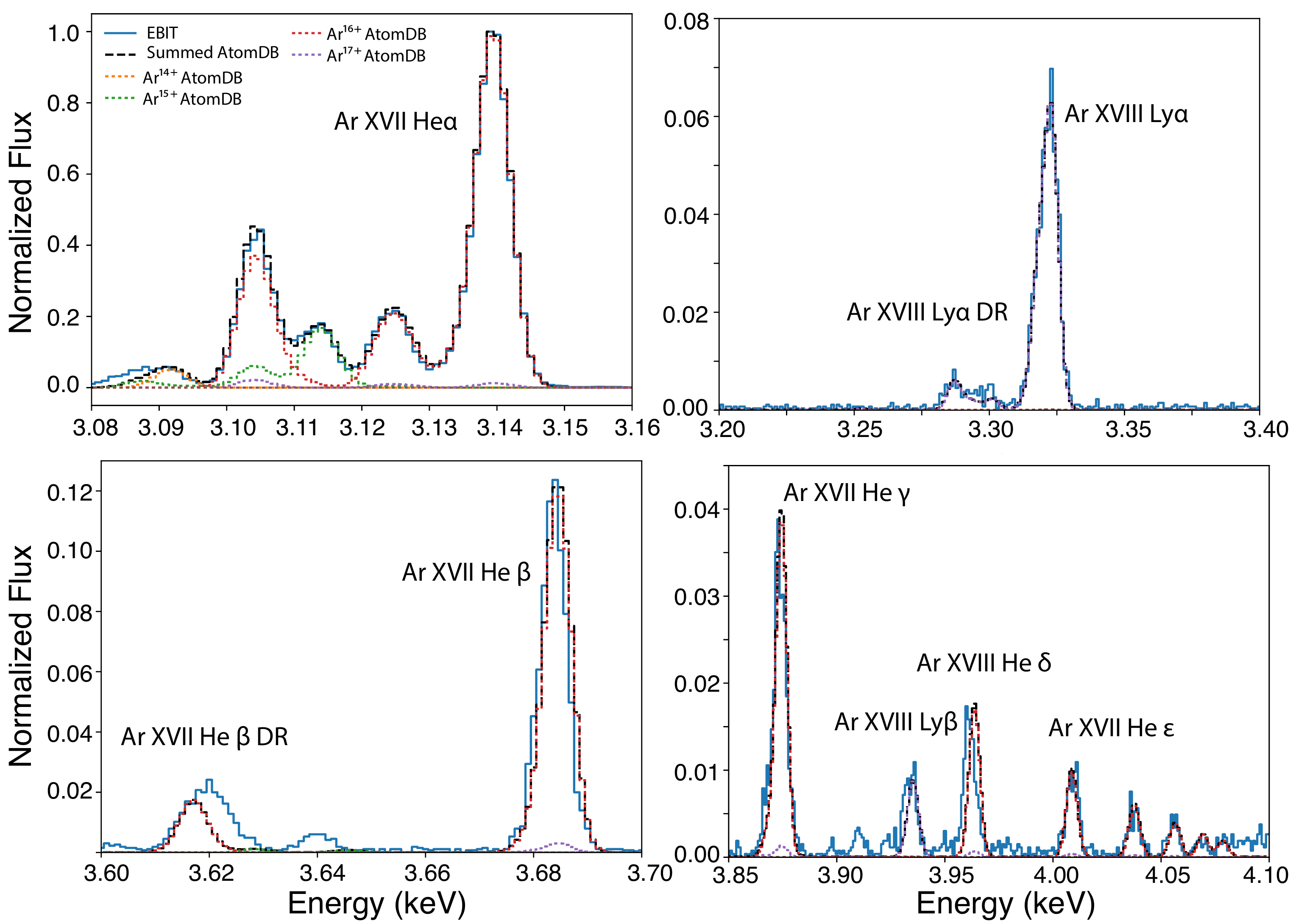}
\caption{Zoomed in energy bands of the \adb\ and EBIT spectra. The figure compares the EBIT results (solid blue line) with the emission from each ion of argon calculated using \adb\ with the ion fractions for the EBIT plasma from Table \ref{tab:ionfrac}.}
\label{fig:maxwebitspec}
\end{figure*}

In Figure \ref{fig:fullspec} we show the spectrum of a 1.74~keV plasma from \adb\ and the observed EBIT spectrum, labeling the strong lines of Ar. 
We stressed that the EBIT spectrum is artificially shifted by +5~eV to show the difference between \adb\ and the measured EBIT spectra. As can be seen, there is a significant difference in the Lyman series line intensities as there are a lot fewer \ionch{Ar}{17} ions in EBIT than predicted for a Maxwellian charge distribution. Figure \ref{fig:maxwebitspec} compares the spectra for four different energy ranges between the EBIT and the spectrum with the fitted EBIT ion ratios (e.g., $3.08-3.16$~keV, $3.2-3.4$~keV, $3.6-3.7$~keV, and $3.85-4.1$~keV). It shows that the model is generally a good fit to the observed data, in particular the Ly$\alpha$ lines as well as the He$\alpha$, He$\beta$, and He$\gamma$. There is a noticeable discrepancy at 3.62~keV, the Ar DR line which we are studying.

For the given charge balance in the EBIT-I Maxwellian plasma, ignoring polarization effects, the ratio  of He-$\beta$ DR to He-$\beta$$_{1,2}$ is $0.36 \pm 0.02$ where the error is given by the quadrature sum of background subtraction and statistics. If we assume maximum polarization for He-$\beta_{1}$ of P=0.63 \citep{smith96} and for the DR satellites of 0.4, the ratio becomes 0.4, i.e., it changes by 10\% \citep{smith00}. However, owing to the fact that the beam dynamics are not well known and the polarization was not measured in-situ, the correction for polarization is not well known. Thus we assume the maximum polarization effect in deriving the uncertainty in these measurements.
\begin{table*}[ht!]
\centering

\caption{Integrated Normalized Flux in the EBIT experiment and modeled plasma.}
\scalebox{0.95}{
{\renewcommand{\arraystretch}{1.15}
{\normalsize \begin{tabular}{lccccccc}
\hline \hline
Line		&	Energy (keV)	& Experiment & \adb & Ratio & FAC\\
\hline
Ar {\sc xvii} He$\alpha$                             &3.080 -- 3.150 & 1.00e+00 & 1.00e+00 &  1.000 & 1.00e+00\\
Ar {\sc xviii} Ly$\alpha$ DR                         &3.260 -- 3.315 & 1.06e-02 & 6.92e-03 &  1.533 & 2.13e-02\\
Ar {\sc xviii} Ly$\alpha$                            &3.315 -- 3.350 & 4.35e-02 & 4.19e-02 &  1.037 & 4.33e-02\\
Ar {\sc xvi} K$\beta$                                &3.539 -- 3.600 & 5.21e-03 & 9.83e-04\footnote{with added DR lines (Section \ref{sec:newdr}), 4.27e-03}&  5.297 & 5.62e-03\\
Ar {\sc xvii} He$\beta$ DR                           &3.600 -- 3.645 & 2.31e-02 & 9.77e-03 &  2.359 & 2.43e-02\\
Ar {\sc xvii} He$\beta$ DR$_1$        &3.600 -- 3.630 & 1.94e-02 & 9.43e-03 &  2.057 & 2.21e-02\\
Ar {\sc xvii} He$\beta$ DR$_2$          &3.630 -- 3.645 & 3.65e-03 & 3.39e-04 & 10.746 & 2.10e-03\\
Ar {\sc xvii} He$\beta$                              &3.670 -- 3.700 & 5.98e-02 & 6.29e-02 &  0.952 & 5.49e-02\\
Ar {\sc xvii} He$\gamma$                             &3.850 -- 3.900 & 2.30e-02 & 2.19e-02 &  1.051 & 2.48e-02\\
Ar {\sc xviii} Ly$\beta$                             &3.925 -- 3.945 & 6.19e-03 & 4.60e-03 &  1.346 & 3.72e-03\\
Ar {\sc xvii} He$\delta$                             &3.950 -- 3.975 & 9.70e-03 & 8.88e-03 &  1.092 & 7.78e-03\\
Ar {\sc xvii} He$\epsilon \rightarrow \iota$         &4.000 -- 4.082 & 1.33e-02 & 1.27e-02 &  1.051 & -\\
\hline\hline
\end{tabular}}}
}
\label{tab:summed}
\vspace{4mm}
\end{table*}
In Table \ref{tab:summed} we list the total normalized flux in each of twelve energy bands from both the theoretical and observed EBIT spectra, each corresponding to a different line. The fluxes are normalized to the \ion{Ar}{17} He-$\alpha$ feature flux for each spectrum. The ratios for the non-DR lines are within 10\% of their predicted values, with the exception of the weak Ly$\beta$ feature at 3.935~keV, which is 35\% stronger in the experimental data. The DR lines for both \ion{Ar}{17} and \ion{Ar}{18} are significantly stronger in the EBIT experiment than predicted by our fit model, by factors of 2.4 and 1.5 respectively. This is traced to missing satellite transitions in each band.

\subsection{Effect of additional DR satellite lines}
\label{sec:newdr}
\begin{figure}
\centering
\includegraphics[width=0.48\textwidth]{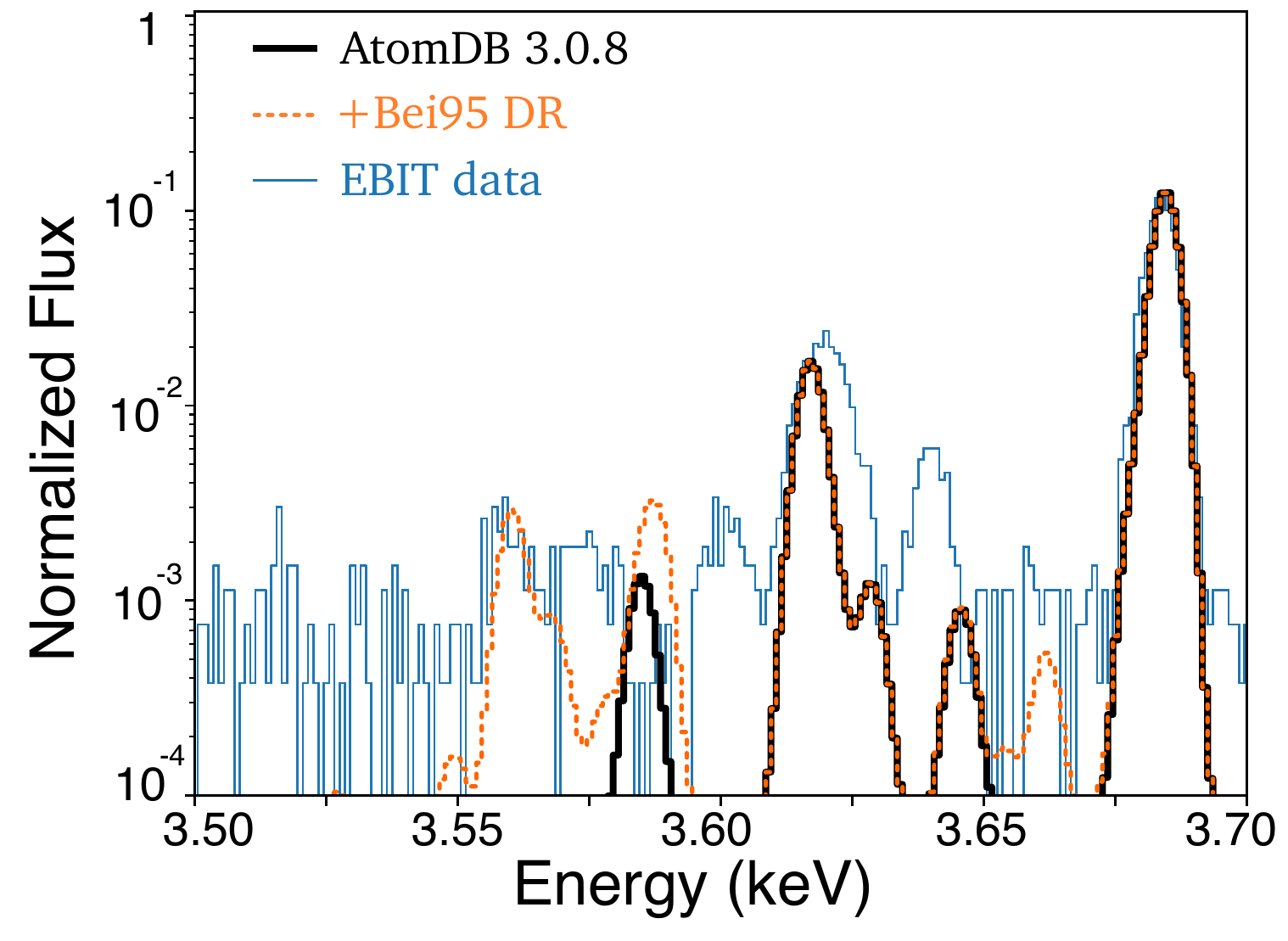}
\caption{The spectrum of a 1.74keV plasma assuming the ion abundances of Table \ref{tab:ionfrac} and \adb\ (solid black line). The orange dotted line shows the same with the DR satellite lines from \cite{bei95} added. EBIT experimental data is shown in blue.}
\label{fig:drsat_bei}
\end{figure}

In this section, we discuss the effects of adding in a range of additional DR satellite lines on our results. In the first case, we added the HULLAC data discussed in Section \ref{sec:AtomicModel:AtomDB}. Including and removing these had no noticeable change on the ion fractions of Table \ref{tab:ionfrac}, with the change in each relative ion fraction being less than 1\%. Where this does make a difference is the \arxvi DR satellites in the  3.55--3.59~keV band. These significantly improve the fit between the EBIT and theoretical data. Figure \ref{fig:drsat_bei} shows  \arxvi XVI DR satellite lines in the 3.5--3.7 keV band, the only region with any significant change. The fluxes of these added lines are still significantly weaker than required to account for the 3.5~keV line flux observed in clusters: they amount to one quarter of the flux in the 3.62~keV line, which is itself too small. Therefore this neither explains the 3.55keV feature in B14a, nor explains the extra flux in the Ar {\sc xvii} He$\beta$ DR lines.

Guided by the lower temperature obtained from the FAC data fit than the AtomDB fit (1.64~keV vs 1.74~keV), we have also investigated a range of electron temperatures in the 1.5 to 1.8~keV range. These yield no significant change ($<2\%$ difference) in the ion fractions, and no significant change in the resulting flux ratios in the 3.62~keV region. Additionally, adjusting the wavelength of the DR satellite lines to match the 3.62~keV line more closely does not lead to any significant change in the modeled ion fractions or line fluxes in this region, as the line's modeled emissivity is constrained by the 3.68~keV line.

\begin{figure*}
\centering
\includegraphics[width=0.99\textwidth]{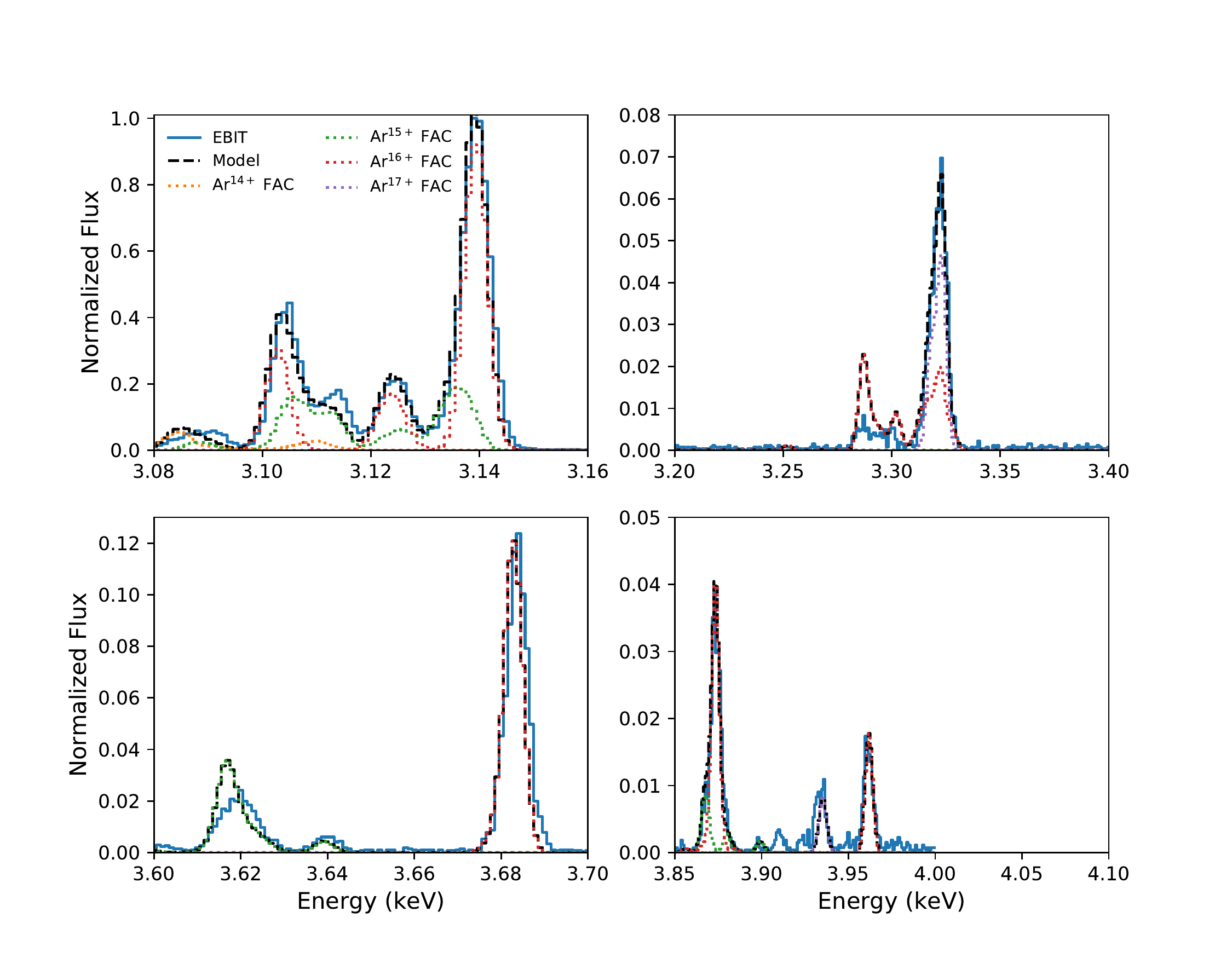}
\caption{Same as Figure \ref{fig:maxwebitspec}, but using FAC to generate the spectrum. The 3.64~keV line is modeled well, though several other features are less well recreated. The spectral model truncates at 4keV.}
\label{fig:facmaxwebitspec}
\end{figure*}

\subsection{Comparison with FAC data}

In an effort to identify the 3.64keV feature, which was not in the AtomDB data, we used the FAC data already calculated to predict this feature, as shown in Figure \ref{fig:facmaxwebitspec}. The results of this suggest the 3.64~keV feature is largely driven by $1s 2p 3[pd] \rightarrow 1s^2 2p$ transitions in the Li like ion. These satellites are not included in the work of \citep{1980ADNDT..25..311V}, which explains their absence from the \adb\ spectrum.

\subsection{Comparison with \citet{b14a}}
One of the suggested interpretations of the 3.5~keV line is an unexpectedly strong \ardr\ dielectronic recombination satellite line created by recombining He-like \arxvii\ at 3.62~keV. Indeed, if the \ardr\ line is $30$ times stronger than the predicted strength in \adbold, this would explain the excess observed in the stacked clusters (see B14a). 
The line ratios and line fluxes presented in this work correspond to a Maxwellian plasma with 1.74~keV temperature that is similar to the intra-cluster medium (ICM) temperatures of cool core clusters.
In this section we fit the \xmm\ spectra of the stacked clusters of galaxies in B14a with the new EBIT measurements to measure the effect of atomic database changes to the detection of the 3.5~keV line.

In the fits of the B14a stacked sample (and in all other samples), the maximum flux for the \ardr\ line at 3.62~keV was initially set to 1\% of the \arxvii\ He$\alpha$ line at 3.12~keV in the spectral fits. The 3.62~keV flux was allowed to go a factor of three above these estimates as safety margin to account for the uncertainties in the emissivities of the weak DR lines. The upper limit corresponds to the highest flux that \ardr\ can have for the lowest ICM plasma temperature (2~keV) observed in B14a.
For instance, the observed flux measured in the \arxvii\ He$\alpha$ line is $2.1\times10^{-5}$ counts cm$^{-2}$ s$^{-1}$. The allowed flux upper limit for the \ardr\ line is $6.3\times10^{-7}$ counts cm$^{-2}$ s$^{-1}$, which is in the limits of the observed ratio by EBIT. Indeed, fitting the 3--6 keV band of the \xmm\ MOS spectrum of the stacked cluster sample with the new ratios indicated by the EBIT-I experiment, the 3.5~keV line is still detected at a 5$\sigma$ confidence level with a flux of 3.98$\pm0.7\ \times10^{-6}$ photons cm$^{-2}$ s$^{-1}$.
Therefore, we rule out here that the excess emission observed in the stacked clusters and the Perseus cluster is due to the satellite \ardr\ lines in the 3.5--3.7~keV energy band.

\section{Conclusions}
\label{sec:conc}
We present lab measurements of the \ionch{Ar}{14} -- \ionch{Ar}{18} spectra using the EBIT-I facility located at LLNL. We obtain a detailed spectrum including faint dielectronic satellites lines and compare the emissivities with the atomic calculations in the atomic database \adb. Our major results are as follows.

\begin{enumerate}
\item We find that the EBIT-I data from a pseudo-Maxwellian plasma of highly charged argon ions are consistent with an ionization temperature of $T_{e}$=~1.7~keV for a plasma with a Maxwellian electron distribution. This temperature is comparable to the plasma temperatures observed in cool-core galaxy clusters, therefore EBIT-I Ar flux measurements can directly be compared to the X-ray data in B14a. 
\item The EBIT-I data are consistent with the predictions of the He-like Ar lines, He$\alpha$, He$\beta$, and He$\gamma$ emissivities of the model within 10\%. 
\item  We have identified new satellite lines in the 3.5--3.7~keV band. These additional lines, however, are too weak to resolve the flux discrepancies near 3.5~keV line as their fluxes are $\sim$0.5\% of that of Ar {\sc XVII} He$\alpha$ line. These have been attributed to the $1s 2p 3[pd] \rightarrow 1s^2 2p$ satellite lines.

\item The measured flux in the \ion{Ar}{17} He-$\beta$ DR line is 2.4 times higher than that predicted in \adb. However, a factor of 3 safety margin was allowed in B14a to account for uncertainties in the satellite line fluxes. Fitting the \xmm\ spectrum of the stacked clusters sample with the ratios indicated by the new EBIT measurements produces a 3.5~keV line flux of 3.98~$\pm\ 0.7\ \times10^{-6}$ photons cm$^{-2}$ s$^{-1}$. The unidentified line is still detected at the 5$\sigma$ confidence level. These new EBIT measurements rule out the \ardr\ origin of the unidentified emission line detected in clusters of galaxies at 3.5~keV.

\end{enumerate}

The EBIT measurements of charged \ionch{Ar}{14} -- \ionch{Ar}{18} ions presented in this work provide independent tests on theoretical calculations and are essential when interpreting the high resolution spectra which will be available through {\it XARM} and {\it Athena} XIFU observations \citep{nandra2013,barret2018}.

%
\section{Acknowledgements}
Authors thank the anonymous referee for helpful comments on the draft. Part of this work was performed under the auspices of the U.S. Department of Energy by Lawrence Livermore National Laboratory under Contract DE-AC52-07NA27344 and supported by NASA contract NNM15AA35I and {\it Chandra} award AR5-16012Z.

\end{document}